# Ground-to-satellite quantum teleportation

Ji-Gang Ren[1,2], Ping Xu[1,2], Hai-Lin Yong[1,2], Liang Zhang[2,3], Sheng-Kai Liao[1,2], Juan Yin[1,2], Wei-Yue Liu[1,2], Wen-Qi Cai[1,2], Meng Yang[1,2], Li Li[1,2], Kui-Xing Yang[1,2], Xuan Han[1,2], Yong-Qiang Yao[4], Ji Li[5], Hai-Yan Wu[5], Song Wan[6], Lei Liu[6], Ding-Quan Liu[3], Yao-Wu Kuang[3], Zhi-Ping He[3], Peng Shang[1,2], Cheng Guo[1,2], Ru-Hua Zheng[7], Kai Tian[8], Zhen-Cai Zhu[6], Nai-Le Liu[1,2], Chao-Yang Lu[1,2], Rong Shu[2,3], Yu-Ao Chen[1,2], Cheng-Zhi Peng[1,2], Jian-Yu Wang[2,3], Jian-Wei Pan[1,2].

[1] Shanghai Branch, Department of Modern Physics and National Laboratory for Physical Sciences at the Microscale, University of Science and Technology of China, Shanghai 201315, China.

[2] CAS Center for Excellence and Synergetic Innovation Center in Quantum Information and Quantum Physics, University of Science and Technology of China, Shanghai 201315, China.

[3] Key Laboratory of Space Active Opto-Electronic Technology, Shanghai Institute of Technical Physics, Chinese Academy of Sciences, Shanghai 200083, China.

[4] National Astronomical Observatories, Chinese Academy of Sciences, Beijing 100012, China.

[5] Nanjing Astronomical Instruments Company Limited, Chinese Academy of Sciences, Nanjing 210042, China.

[6] Shanghai Engineering Center for Microsatellites, Shanghai 201203, China.

[7] Beijing Institute of Tracking and Telecommunication Technology, Beijing 100094, China.

[8] State Key Laboratory of Astronautic Dynamics, Affiliated to China Xi'an Satellite Control Center , 710061, China.

An arbitrary unknown quantum state cannot be precisely measured or perfectly replicated[1]. However, quantum teleportation allows faithful transfer of unknown quantum states from one object to another over long distance[2], without physical travelling of the object itself. Long-distance teleportation has been recognized as a fundamental element in protocols such as large-scale quantum networks[3,4] and distributed quantum computation[5,6]. However, the previous teleportation experiments between distant locations[7-12] were limited to a distance on the order of 100 kilometers, due to photon loss in optical fibres or terrestrial free-space channels. An outstanding open challenge for a global-scale "quantum internet"[13] is to significantly extend the range for teleportation. A promising solution to this problem is exploiting satellite platform and space-based link, which can conveniently connect two remote points on the Earth with greatly reduced channel loss because most of the photons' propagation path is in empty space. Here, we report the first quantum teleportation of independent single-photon qubits from a ground observatory to a low Earth orbit satellite—through an up-link channel—with a distance up to 1400 km. To optimize the link efficiency and overcome the atmospheric turbulence in the up-link, a series of techniques are developed, including a compact ultra-bright source of multi-photon entanglement, narrow beam divergence, high-bandwidth and high-accuracy acquiring, pointing, and tracking (APT). We demonstrate successful quantum teleportation for six input states in mutually unbiased bases with an average fidelity of 0.80±0.01, well above the classical limit[14]. This work establishes the first ground-to-satellite up-link for faithful and ultra-long-distance quantum teleportation, an essential step toward global-scale quantum internet.

In our experiment, the quantum state to be teleported is the polarization of a single photon which can be written as: $|\chi\rangle_1 = \alpha|H\rangle_1 + \beta|V\rangle_1$, where $\alpha$ and $\beta$ are two unknown, complex numbers satisfying $|\alpha|^2 + |\beta|^2 = 1$. $|H\rangle$ and $|V\rangle$ denote the horizontal and vertical polarization states, respectively, and can be used to encode the basic logic $|0\rangle$ and $|1\rangle$ for a quantum bit (qubit). Such a single qubit is generated from an observatory ground station in Ngari (located at Tibet; latitude 32°38'43.07" N, longitude 98°34'18.80" E; altitude 5100 m), and aimed to be teleported to the *Micius* satellite that has been launched from China on 16th August 2016 to an altitude of ~500 km. The satellite flies along a sun-synchronous orbit, i.e., it passes over any given point of the planet's surface at the same local solar time (00:00 midnight).

Quantum teleportation[2], proposed by Bennett *et al.*, relies on using both a classical channel and a quantum channel—entanglement—that are shared between the two communicating parties, whom we called as Ngari and *Micius*. The entangled state of a pair of photons can be written as $|\phi^+\rangle_{23} = (|H\rangle_2|H\rangle_3 + |V\rangle_2|V\rangle_3)/\sqrt{2}$, one of the four maximally entangled two-qubit Bell states. Ngari performs a joint measurement on the to-be-teleported photon 1 and the photon 2 from the entangled pair, projecting them into one of the four Bell states. In the case that the Bell-state measurement (BSM) yields $|\phi^+\rangle_{12}$, the photon 3 carries exactly the desired state. If another Bell state, $|\phi^-\rangle_{12} = (|H\rangle_1|H\rangle_2 - |V\rangle_1|V\rangle_2)/\sqrt{2}$ is detected, then up to a unitary $\pi$ phase shift in the data post-processing, the state of photon 3 is equivalent to the original state of the photon 1.

Experimentally, the realization of quantum teleportation of an independent single photon necessitates the simultaneous creation of two entangled photon pairs[15] and high-visibility quantum interference between them[7]. The multi-photon coincidence count rate is several orders of magnitude lower compared to typical single- or two-photon experiments. Due to the complexity of the multi-photon set-up for space-scale quantum teleportation, we choose the uplink configuration (Fig. 1a) where the transmitter (Ngari) is placed in the ground station and the receiver (*Micius*) is in the satellite. To maximize the experiment count rate, we prepare compact and ultra-bright four-photon sources (see Fig. 1b for a schematic drawing). Ultraviolet femtosecond pulses (with a central wavelength of 390 nm, a pulse width of 160 fs, and a repetition rate of 80 MHz) from a mode-locked Ti:sapphire laser passes through two bismuth borate (BiBO) crystals to generate two pairs of photons. The first pair is generated via collinear SPDC where one of them is detected as a trigger to herald the presence of a single photon 1 (count rate $5.7 \times 10^5$/s), whose state is to be teleported. Using a half-wave plate (HWP) and a quarter-wave plate (QWP), the initial polarization state of the photon 1 can be arbitrarily prepared. The second BiBO crystal was aligned for non-collinear SPDC[16] and prepared in the frequency-uncorrelated polarization-entangled state $|\phi^+\rangle_{23}$, with a count rate of $1 \times 10^6$/s and a fidelity of 0.933 measured on the ground (see Methods).

To realize the BSM, the photon 1 and 2 are then overlapped on a polarizing beam splitter (PBS), which transmit *H* and reflect *V* polarization. After the PBS, we select these events that each output detects one photon, which is possible only if the two photons have the same polarization— $|H\rangle_1|H\rangle_2$ or $|V\rangle_1|V\rangle_2$ —thus projecting the wave function into a subspace of $|\phi^\pm\rangle_{12} = (|H\rangle_1|H\rangle_2 \pm |V\rangle_1|V\rangle_2)/\sqrt{2}$. Finally, by measuring the photons in the $(|H\rangle \pm |V\rangle)/\sqrt{2}$ bases at the outputs of the PBS, we can distinguish $|\phi^+\rangle_{12}$ and $|\phi^-\rangle_{12}$ (ref. 17). After the BSM, the final four-photon count rate measured on the ground is 4080/s. To achieve a high stability, the four-photon interferometry system is integrated into a compact platform with a dimension of $460\,\text{mm} \times 510\,\text{mm} \times 100\,\text{mm}$ and a weight of less than 20 kg (see Methods and Extended Data Fig. 2). The variation of the four-photon count rate is observed to be less than 10 % for a duration of two weeks when the setup is mounted in Ngari observatory station. Using the same pump laser, a second multi-photon module with the same design is built in sequence, which increases the four-photon count rate to 8210 /s by multiplexing (see Methods).

Compared to our downlink experiment[18], a significant challenge of the uplink channel in the present work is that the atmospheric turbulence occurs at the beginning of the transmission path, which causes beam wandering and broadening that increases the amount of spreading of the travelling beams. We design a transmitting telescope with narrow divergence, and develop a high-bandwidth and high-precision APT system to optimize the uplink efficiency. The multi-stage APT system consists of both coarse and fine tracking with a tracking error of ~3 μrad (see Methods and Extended Data Fig. 4). The teleported single photons from a single-mode fibre are transmitted through a 130-mm-diameter off-axis reflecting telescope (Fig. 1c), and received by a

300-mm-diameter telescope equipped in the satellite (Fig. 1d). The locally tested beam divergence angle of the transmitting antenna is ~14 ± 1 μrad (see Extended Data Fig. 4c), measured with a long focal length collimator on the ground. The atmospheric seeing in Ngari is on the order of 5 μrad, which will in principle increase the divergence angle to ~15 μrad. In our experiment, we couple the photons emitted from stars into a single-mode fibre, and measure the intensity distribution as a function of the fine-tracking scanning angle. The effective divergence angle estimated from the measured field of view (FOV) of the intensity distribution (see Extended Data Fig. 4d) is 22 ± 3 μrad, because many additional factors can make the scan result of the FOV larger, including the mismatch of the diffraction spot size and the fiber core radius, the altitude angle of star, the precision of the tracking, the changes in atmospheric environment. Finally, the beam divergence involving the fast-flying satellite is tested by sending an attenuated laser (~20 billion photon per second) to the satellite, which is collected by the satellite by varying the fine-tracking angle. The obtained intensity pattern is elliptical with an equivalent divergence of 24-35 μrad. Figure 2 shows a time-trace of channel attenuation measured during one orbit of the satellite passing through the Ngari station. The physical distance from the ground station and the satellite varies from a maximal of 1400 km (at an altitude angle of 14.5°, the starting point of our measurement) to a minimal of 500 km (at the highest altitude angle of 76.0°, when the satellite passes through the ground station above the top), where the channel loss of the uplink is from 52 dB to 41 dB measured using a high-intensity reference laser.

After passing through the uplink, the teleported photon is detected by two fibre coupled silicon avalanche photodiodes (Fig. 1d). Before entering in orbit, the dark count rate of the detectors is ~20 Hz. As the detectors are exposed to radiation in the space environment, they are carefully shielded and cooled down to -50℃ to reduce the dark counts to less than 150 Hz over a 3 month period. A 3-nm narrowband filter is placed before the detectors to block stray light from the reflection of the moonlight (maximal ~350 Hz at full moon). Further, time synchronization between the satellite and ground is employed (see Methods) to reliably extract four-photon coincidence counts within a time window of 3 ns.

To demonstrate that the quantum teleportation is universal, we test six input states in mutually unbiased bases on the Bloch sphere: $|H\rangle_1$, $|V\rangle_1$, $|+\rangle_1 = (|H\rangle_1 + |V\rangle_1)/\sqrt{2}$, $|-\rangle_1 = (|H\rangle_1 - |V\rangle_1)/\sqrt{2}$, $|R\rangle_1 = (|H\rangle_1 + i|V\rangle_1)/\sqrt{2}$, and $|L\rangle_1 = (|H\rangle_1 - i|V\rangle_1)/\sqrt{2}$.

To evaluate the performance of the teleportation, we measure the teleported state fidelity $F = Tr(\hat{\rho}|\chi\rangle\langle\chi|)$, defined as the overlap of the ideal teleported state ($|\chi\rangle$) and the measured density matrix ($\hat{\rho}$). The teleported photon 3 is measured using a polarization analyser that comprises of a QWP, a HWP (both installed inside remotely controlled rotation mounts), and a PBS, followed by two single-photon detectors, which projects the photon 3 either to the ideal state $|\chi\rangle$ or its orthogonal state $|\chi\rangle^\perp$. Conditioned on the ground detection of the trigger photon and the two-photon double click after the BSM, we register the photon counts of the teleported photon 3 using the two-channel polarization analyser on the satellite, and record the two sets of data.

After applying 0 or π phase shift in the data post-processing, depending on the respective outcome state $|\phi^+\rangle_{12}$ or $|\phi^-\rangle_{12}$ of the BSM, the teleportation fidelity can be calculated by the ratio of the correct four-photon coincidence counts to the overall four-photon events. We obtain overall 911 four-photon counts in 32 orbits, each orbit taking 350 s for data collection. In the 32 different days, the orbits vary and thus as the shortest distances between the satellite to the ground station (see Extended Data Table II for a summary). For the set of the six input states, the teleportation state fidelities are summarized in Fig. 3, yielding an average $\overline{F} = 0.80 \pm 0.01$, sampling over the whole Bloch sphere. We note that all reported data are without background subtraction.

The main sources of fidelity error include double pair emission of SPDC (6%), partial photon distinguishability (10%), uplink polarization distortion (3%), and background dark count (4%). See Methods for a more detailed analysis. Despite the photon loss and environmental noise, the measured teleportation state fidelities are all well above 2/3 —the classical limit, defined as the optimal state-estimation fidelity on a single copy of a single qubit system[14] that one can reach with a classical strategy without sharing entanglement as a resource. These results conclusively confirm the quantum nature of teleportation of a single qubit.

In summary, our work has established the first ground-to-satellite uplink at ~500-1400 kilometre scale with 41-52 dB loss, and accomplishes the faithful transfer of the superposition state of a single-photon qubit using the quantum teleportation. As a comparison, if one employs the same four-photon source and sends the teleported photon through an 1200-km telecommunication fibre with 0.2 dB/km loss, it is straightforward to check that one would have to wait for 380 billion years (20 times the Universe's lifetime) to witness one event, assuming the detectors have absolutely no dark counts.

In the current work, the entangled photon source and the BSM are performed at the same location on the ground. A next step toward real network connections is to realize long-distance entanglement distribution prior to the BSM[10-12]. To this aim, one approach is to develop entangled-photon source with long coherence time ($T_c$) and to reduce the arrival time jitter ($T_j$) between independent photons such that $T_c > T_j$. Teleportation is not restricted to photons as in this work, but would also allow, for example, transferring the quantum state of a fast-flying single photon into a long-lived matter qubit as a quantum memory at a distance[19-21]. Teleportation of a subsystem of an entangled pair translates itself into the protocol of entanglement swapping[22-25] where two remote particles can become entangled without direct interactions. Further, teleportation of quantum logic gates, a key element in distributed quantum computing schemes, is also possible assisted by shared multiparticle entangled states[5,6]. Given the rapid progress on long-live quantum memories[26] and efficient light-matter interface[27], more sophisticated space-scale teleportation can be realized and is expected to play a key role in the future distributed quantum internet.

matter interface with sub-second lifetime. *Nat. Photon.* **10**, 381-384 (2016).

**Figure caption:**

Figure 1: Overview of the set-up for ground-to-satellite quantum teleportation of a single photon with a distance up to 1400 km. **a**, The background is a photograph of the Ngari ground station in Tibet. **b,** The compact multi-photon set-up on the ground for teleportation. The 390 nm pulsed laser passes through two closely mounted BiBO crystals, and produces two photons pairs through collinear SPDC (the two photons are in the same spatial mode 1') and non-collinear SPDC (the two photons are separated in the spatial modes 2' and 3'). The photon 2' and 3' are then superposed on a PBS to disentangle their frequency information from polarization information for the preparation of entangled photons (labelled as 2 and 3) with both high brightness and high fidelity[16]. The two photons from collinear SPDC in the spatial mode 1' are first separated from the pump beam using a dichroic mirror (DM), and then separated by a PBS where the state of the transmitted photon (labelled as 1) is to be teleported whereas the reflected photon serves as a trigger. Photon 1 and 2 are then combined on another PBS for Bell-state measurement. Narrowband filters with 3 nm and 8 nm are used to erase the frequency correlation of the down-converted photons. The teleported photon 1 is collected by a single-mode fibre and guided to c. To increase the overall four-photon count rate, a second four-photon module (not shown) with the same design and using the same pump laser is prepared and used for multiplexing (see Methods). **c**, The transmitting antenna. A faster-steering mirror (FSM) and two-axis gimbal mirrors are used for fine and coarse tracking. The DM is used to separate the signal and beacon light. The top right panel illustrates a 671 nm cw laser and a 1064 nm pulsed laser with a repetition rate of 10-kHz shooting from ground to the satellite for APT and time synchronization, respectively. A CMOS camera is used to imaging the beacon laser from the satellite. **d,** The receiver on the satellite. A similar APT system is implemented as the ground station. A polarization analysis consisting of a QWP, a HWP, and a PBS is used, followed by two single-photon detectors.

**Figure 2**: Distance from the ground station to the orbiting satellite and the measured attenuation during one orbit. **a,** The trajectory of the *Micuis* satellite measured from Ngari ground station in one orbit with a duration of 350 s. **b,** The measured ground-to-satellite channel loss using a strong reference laser as a function of passing time. The highest loss is ~52 dB at a distance of 1400 km, when the satellite is at 14.5° angle. The lowest loss is ~41 dB at a distance of ~500 km, when the satellite is at 76.0° angle. The error bars represent 1 s.d., deduced from propagated poissonian counting statistics of the raw detection events. The red curve is a supposed model (see Methods) which considers the effect of distance variation. However, due to the structure of the altazimuth telescope in the ground station, the rotation speed of the optical transmitter has to be increased as a function of the increasing rising altitude angle of the satellite. When the satellite reaches the top altitude angle, the needed speed can be very large and out of the reach of the APT system ability. Therefore, with the increase of the rotational speed, the tracking accuracy was reduced. This leads to the measured channel attenuation larger when the satellite is closer to the ground station.

**Figure 3**: Teleportation state fidelities for the six quantum states with data taken for 32 orbits. The details of the date, the highest altitude angles, and the ground-to-satellite distance of the 32 orbits are listed in Extended Data Table II. All the fidelities are well above the classical limit of 2/3 shown as the dashed line. The error bars represent one standard deviation calculated from Poissonian counting statistics of the raw detection events.

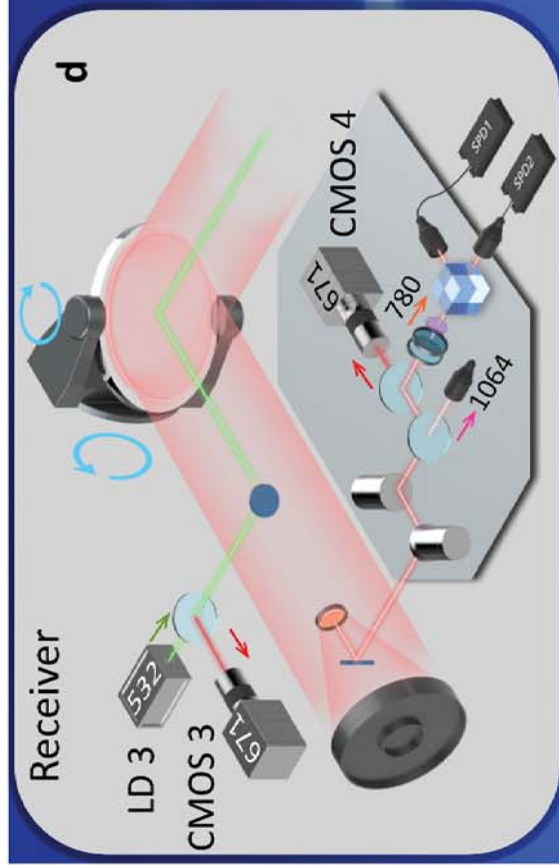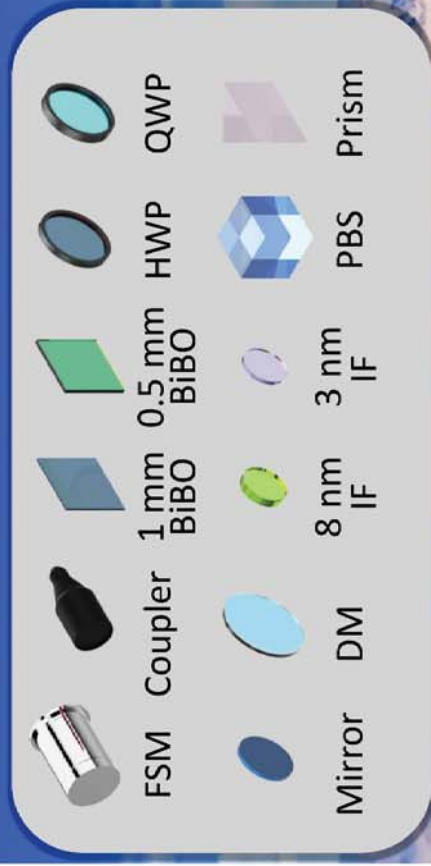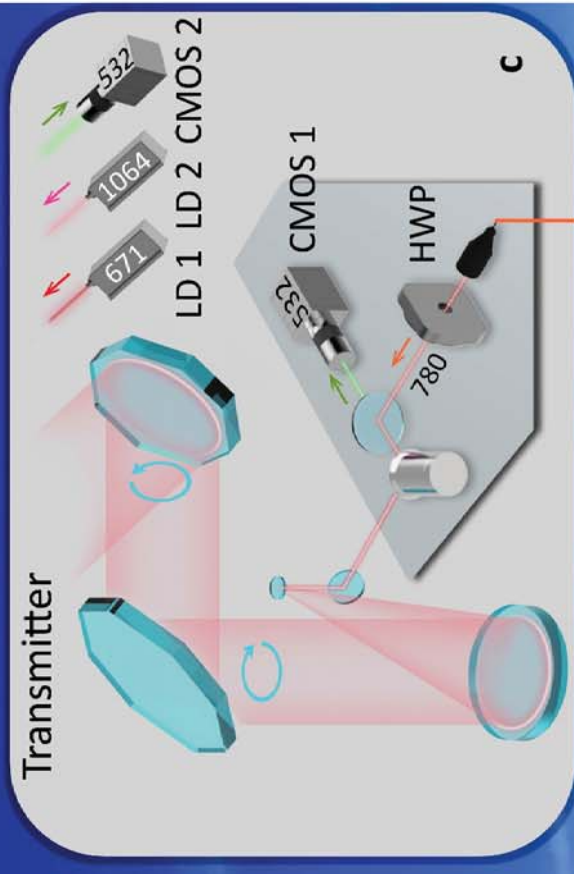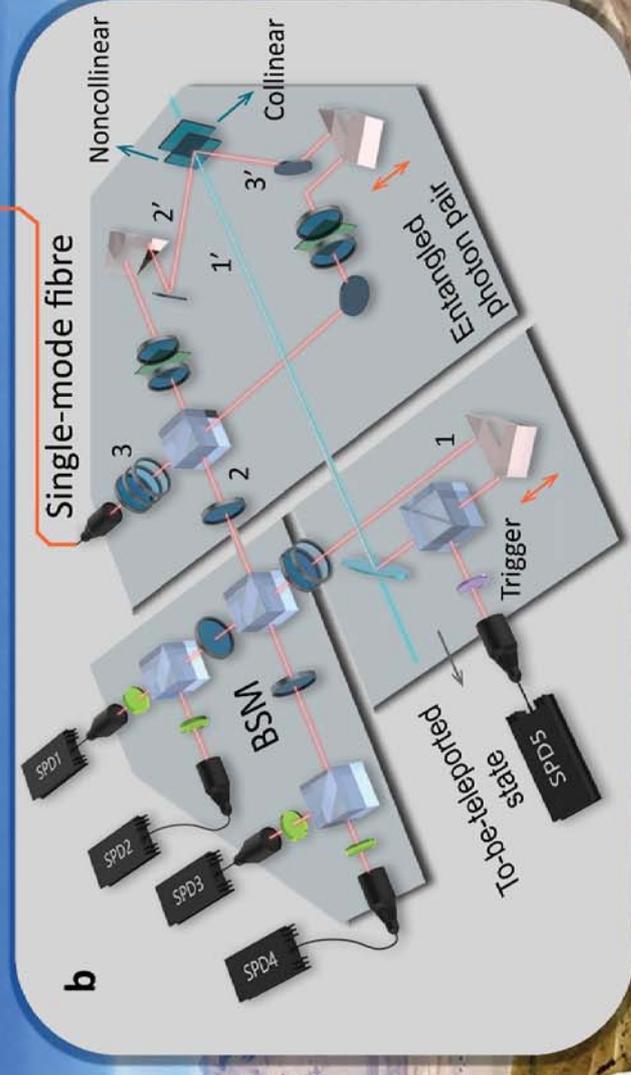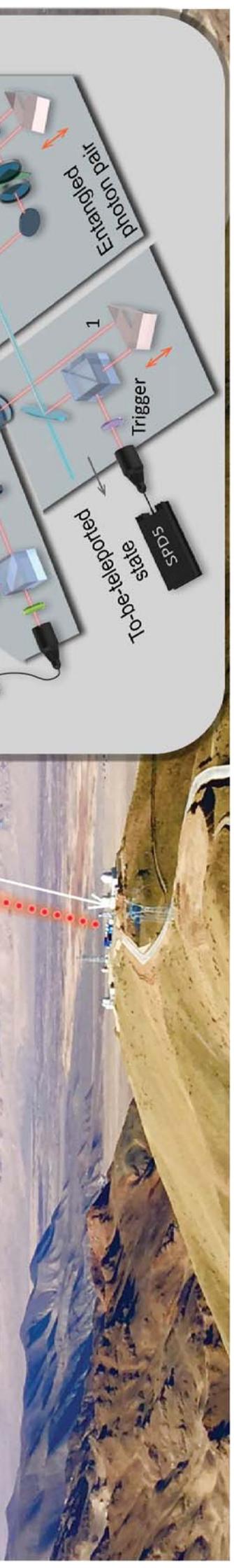

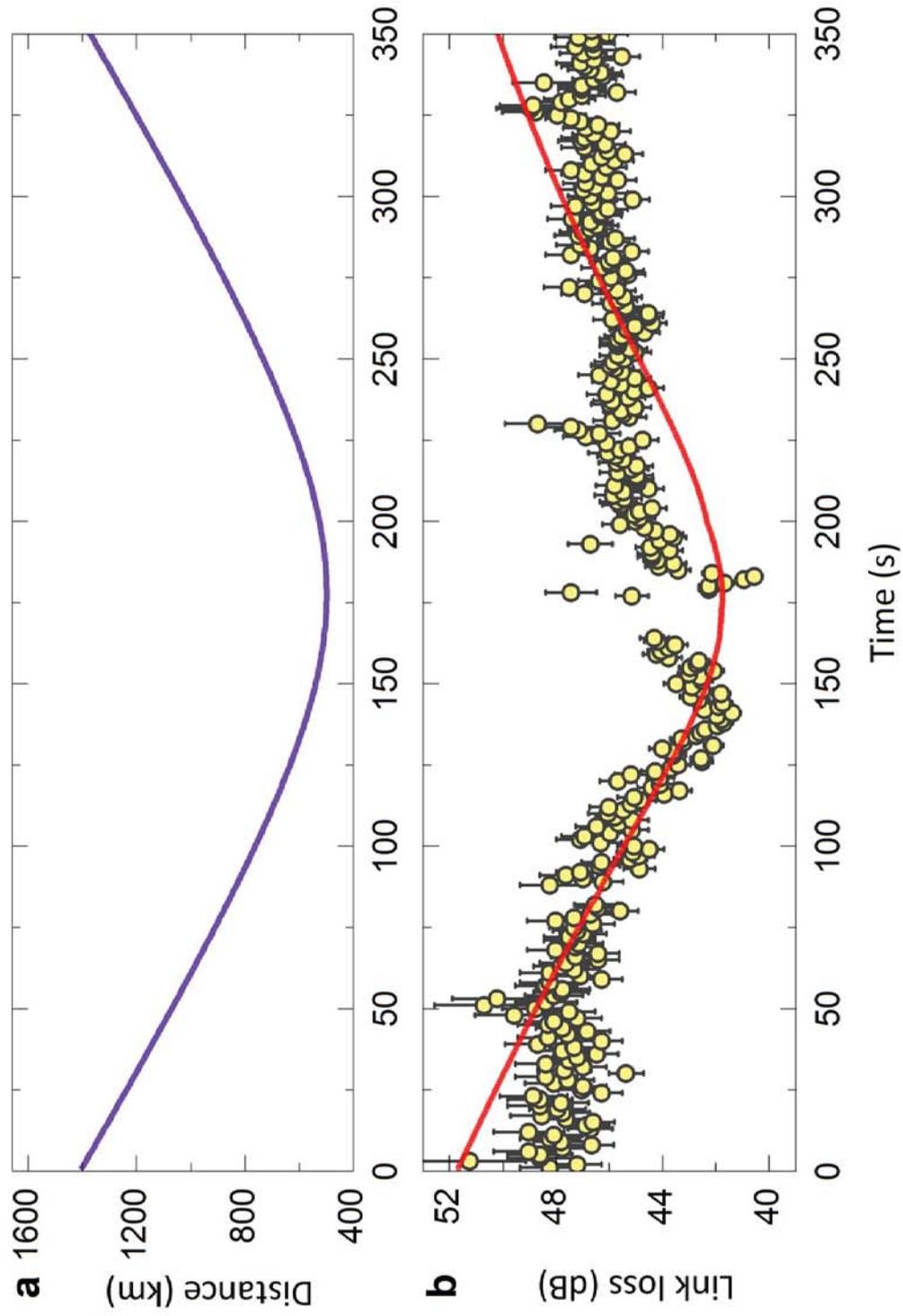

Fig. 2

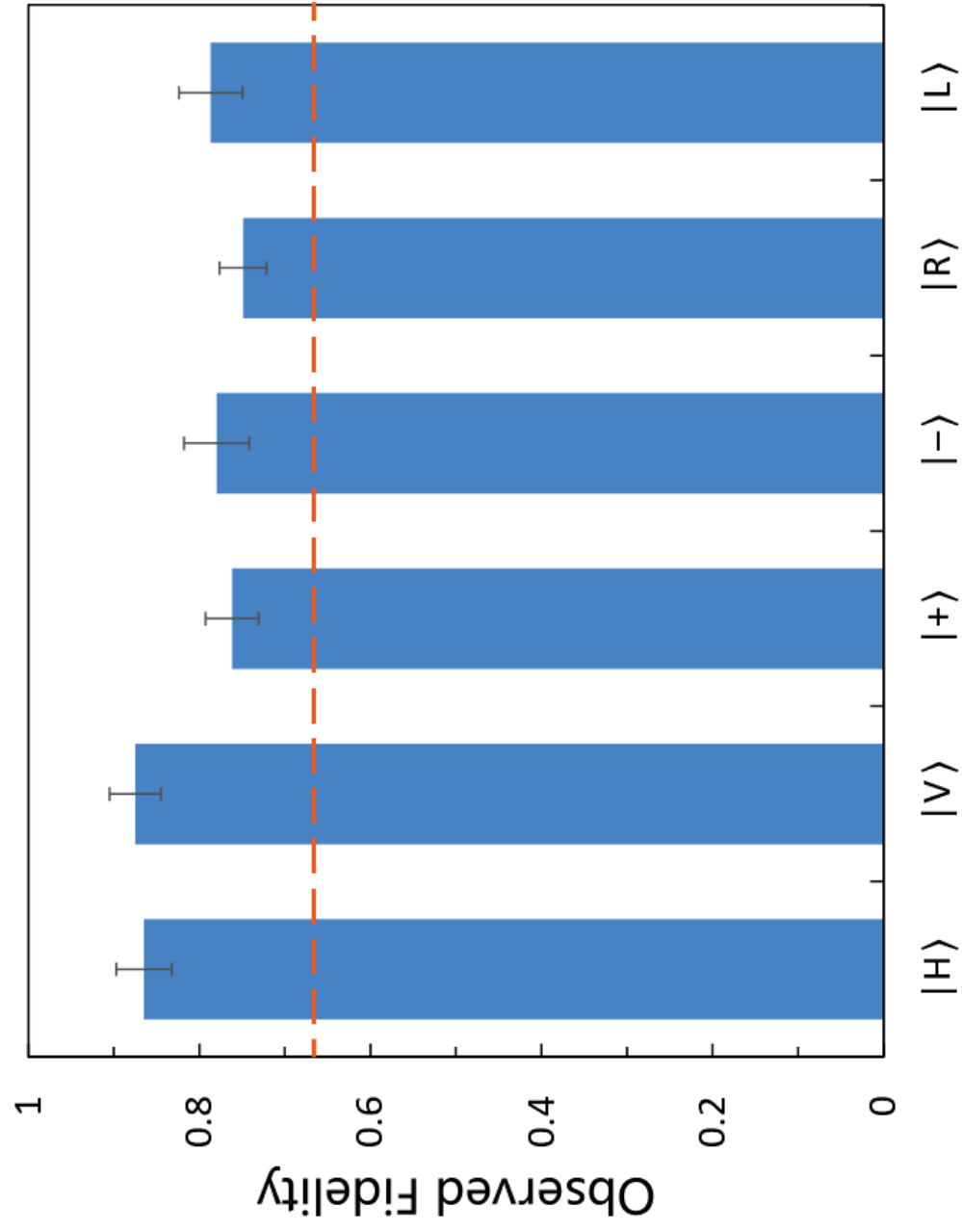

Fig. 3